\documentclass[sigconf]{acmart}
\usepackage{pbalance}
\usepackage{enumitem}
\usepackage{booktabs}
\usepackage{graphicx}
\usepackage{array}
\usepackage{lineno}
\usepackage{multirow}
\usepackage{hhline}
\usepackage{makecell}
\usepackage{subfigure}
\usepackage{bbm}
\usepackage[marginal]{footmisc}

\AtBeginDocument{%
  \providecommand\BibTeX{{%
    \normalfont B\kern-0.5em{\scshape i\kern-0.25em b}\kern-0.8em\TeX}}}

\copyrightyear{2026}
\acmYear{2026}
\setcopyright{cc}
\setcctype{by-nc-nd}
\acmConference[RecSys '26]{20th ACM Conference on Recommender Systems}{September 27-October 02, 2026}{Minneapolis, MN, USA}
\acmBooktitle{20th ACM Conference on Recommender Systems (RecSys '26), September 27-October 02, 2026, Minneapolis, MN, USA}
\acmDOI{10.1145/3773078.3831746}
\acmISBN{979-8-4007-2284-4/2026/09}

\begin{document}

\title{
Sample Is Feature: Beyond Item-Level, Toward Sample-Level Tokens for Unified Large Recommender Models
}


\author{Shuli Wang}
\authornote{Corresponding author.}
\affiliation{%
  \institution{Meituan}
   \city{Chengdu}
  \country{China}
}
\email{wangshuli03@meituan.com}

\author{Junwei Yin}
\affiliation{%
\institution{Meituan}
   \city{Chengdu}
  \country{China}
  }
\email{yinjunwei03@meituan.com}

\author{Changhao Li}
\affiliation{%
  \institution{Meituan}
   \city{Chengdu}
  \country{China}
}
\email{lichanghao@meituan.com}

\author{Senjie Kou}
\affiliation{%
\institution{Meituan}
   \city{Chengdu}
  \country{China}
  }
\email{kousenjie@meituan.com}

\author{Chi Wang}
\affiliation{%
\institution{Meituan}
   \city{Chengdu}
  \country{China}
  }
\email{wangchi06@meituan.com}

\author{Yinqiu Huang}
\affiliation{%
\institution{Meituan}
   \city{Chengdu}
  \country{China}
  }
\email{huangyinqiu@meituan.com}

\author{Yinhua Zhu}
\affiliation{%
\institution{Meituan}
   \city{Chengdu}
  \country{China}
  }
\email{zhuyinhua@meituan.com}

\author{Haitao Wang}
\affiliation{%
\institution{Meituan}
   \city{Chengdu}
  \country{China}
  }
\email{wanghaitao13@meituan.com}

\author{Xingxing Wang}
\affiliation{%
\institution{Meituan}
   \city{Beijing}
  \country{China}
  }
\email{wangxingxing04@meituan.com}

\renewcommand{\shortauthors}{Wang et al.}




\begin{abstract}
Scaling industrial recommender models has followed two parallel paradigms: \textbf{sample information scaling}---enriching the information content of each training sample through deeper and longer behavior sequences---and \textbf{model capacity scaling}---unifying sequence modeling and feature interaction within a single Transformer backbone. However, these two paradigms still face two structural limitations. Firstly, sample information scaling methods encode only a subset of each historical interaction into the sequence token, leaving the majority of the original sample context unexploited and precluding the modeling of sample-level, time-varying features. Secondly, model capacity scaling methods are inherently constrained by the structural heterogeneity between sequential and non-sequential features, preventing the model from fully realizing its representational capacity.

To address these issues, we propose \textbf{SIF} (\emph{Sample Is Feature}), which encodes each historical Raw Sample directly into the sequence token---maximally preserving sample information while simultaneously resolving the heterogeneity between sequential and non-sequential features. SIF consists of two key components. The \textbf{Sample Tokenizer} quantizes each historical Raw Sample into a Token Sample via hierarchical group-adaptive quantization (HGAQ), enabling full sample-level context to be incorporated into the sequence efficiently. The \textbf{SIF-Mixer} then performs deep feature interaction over the homogeneous sample representations via token-level and sample-level mixing, fully unleashing the model's representational capacity. Extensive experiments on a large-scale industrial dataset validate SIF's effectiveness, and we have successfully deployed SIF on an industrial food delivery platform.
\end{abstract}

\ccsdesc[500]{Information systems~Recommender systems}

\keywords{Recommender Systems,  Ranking Model, Scaling Laws}

\maketitle

\section{Introduction}

Recommender systems are a core infrastructure for large-scale internet platforms, connecting hundreds of millions of users to relevant content across e-commerce, short video, and local services. In industrial food delivery platforms, hundreds of billions of daily recommendation requests are served, where even marginal improvements in ranking quality translate directly into business value. With the advancement of large-scale neural models, scaling---continuously improving model performance by expanding training data, sequence length, or model capacity---has become one of the most active research directions in industrial recommender systems.

Existing scaling efforts fall into two complementary paradigms. The first, \textbf{sample information scaling}, enriches the information content of each training sample by extending and deepening historical behavior sequences. This has been pursued along two sub-directions: \emph{sequence lengthening}, which extends behavior history from tens to thousands of interactions (SIM~\cite{sim2020}, ETA~\cite{eta2020}, TWIN~\cite{twin2023}, LONGER~\cite{longer}), and \emph{sequence widening}, which enriches each historical token with additional context beyond the bare item embedding (DSAN~\cite{dsan}, CAIN \cite{cain}, HSTU~\cite{hstu2024}). The second, \textbf{model capacity scaling}, enlarges model expressiveness by unifying sequence modeling and feature interaction within a single Transformer backbone. Rather than processing the behavior sequence and feature interactions in separate modules, methods such as HyFormer~\cite{hyformer2024}, OneTrans~\cite{onetrans2023}, and MixFormer~\cite{mixformer} feed all feature fields---including historical tokens---into a unified deep architecture, enabling richer cross-field interactions in a single forward pass.

However, despite their respective advances, these two paradigms still face two structural limitations. Firstly, sample information scaling methods encode only a subset of each historical interaction into the sequence token: due to storage and serving cost constraints, they either retain bare item embeddings or enrich tokens with only selected feature subsets, leaving the majority of the original sample context---user profile, contextual signals, behavioral outcome---unexploited. More critically, this architectural choice renders these methods structurally incapable of modeling \emph{sample-level, time-varying features}---signals such as real-time item popularity, competitive exposure, and temporal demand shifts that are unique to each individual impression and cannot be recovered from any static item representation. Secondly, model capacity scaling methods are inherently constrained by the structural heterogeneity between sequential and non-sequential features: item-level history tokens and rich multi-field current-request tokens differ fundamentally in information density, and placing them in the same attention space creates a representational asymmetry that prevents the model from fully realizing its capacity.

Consider two users who both clicked the same restaurant: one during a late-night session with a discount coupon active, the other during a lunch rush on a weekday. A bare item embedding treats these interactions as identical, yet their full request contexts carry very different signals about user intent. Preserving the complete Raw Sample for each interaction---rather than a stripped item embedding---could substantially improve the model's ability to reason about temporal and contextual patterns.

To address the aforementioned challenges, we propose \textbf{SIF} (\emph{Sample Is Feature}), which encodes each historical Raw Sample directly into the sequence token---maximally preserving sample information while simultaneously resolving the heterogeneity between sequential and non-sequential features. The key insight is that, at training time, every historical interaction already has an associated full request record---user profile, item features, contextual signals, and pre-computed cross features---stored in the training log; the bottleneck lies not in data availability---every historical interaction is already logged---but in representation: the rich sample context stored in training logs has not been fully exploited. SIF closes this gap with two key components. The \textbf{Sample Tokenizer} compresses each Raw Sample offline into a Token Sample via hierarchical group-adaptive quantization (HGAQ), jointly optimized with the ranking objective, so that full sample-level context can be retrieved at serving time with negligible overhead. The resulting sequence consists entirely of structurally homogeneous Token Samples, eliminating the representational asymmetry between historical and current-request features. The \textbf{SIF-Mixer} then performs deep feature interaction over these homogeneous representations via token-level and sample-level mixing, enabling the model to fully exploit its representational capacity across both intra-sample and inter-sample dimensions.

In summary, our contributions are as follows:
\begin{itemize}
  \item We propose \textbf{SIF}, a framework that elevates each historical sequence token from item-level to sample-level, simultaneously resolving the incomplete sample utilization in sample information scaling and the feature heterogeneity in model capacity scaling.
  \item We introduce the \textbf{Sample Tokenizer}, which compresses Raw Samples into Token Samples via hierarchical group-adaptive quantization (HGAQ) for efficient serving, and the \textbf{SIF-Mixer}, which performs token-level and sample-level feature interaction over the resulting homogeneous sample representations.
  \item We conduct extensive experiments on a large-scale industrial dataset, demonstrating that SIF consistently outperforms strong baselines with up to \textbf{+0.88\% GAUC} offline, and achieves \textbf{+2.03\% CTR} and \textbf{+1.21\% CVR} in online A/B testing on an industrial food delivery platform.
\end{itemize}

\section{Related Work}

\subsection{Sample Information Scaling}

Sample Information Scaling improves recommendation quality by enriching the information carried per historical interaction. It has two complementary sub-directions: \emph{sequence lengthening} (more tokens) and \emph{sequence widening} (richer tokens).

\textbf{Sequence lengthening} extends behavior history from tens to thousands of interactions. DIN~\cite{din2018} introduced attention-based interest extraction; DIEN~\cite{dien2019} modeled interest evolution via GRU. SIM~\cite{sim2020} and ETA~\cite{eta2020} scaled to ultra-long sequences via two-stage retrieval, and LONGER~\cite{longer} generalized this with hierarchical compression. Despite growing sequence lengths, all these methods leave each historical token as a bare item embedding---intra-token information density is an untouched bottleneck.

\textbf{Sequence widening} enriches each historical token with additional context beyond the item embedding. DSAN~\cite{dsan} appends context features to sequence tokens. TWIN~\cite{twin2023} concatenates \emph{target-item} features to each history token to enable target-aware retrieval. SINE~\cite{sine2021} incorporates label and intent signals via contrastive learning. Most relevant to SIF is HSTU~\cite{hstu2024}, which encodes rich per-interaction side information (item features, engagement signals, and context) at Meta's scale via linear attention over thousands of interactions. However, none of these methods can fully preserve the original sample information in each historical token: due to storage and serving cost constraints, they enrich tokens with only \emph{selected} feature subsets---direct concatenation of all feature fields would multiply the sequence token dimension by an order of magnitude, making attention prohibitively expensive at $L = 200$, and storing full float16 snapshots at scale incurs prohibitive storage cost that significantly limits practical applicability.

SIF completes the sequence widening direction by compressing the complete Raw Sample---rather than selected feature subsets---offline via group-wise RVQ; the per-token serving cost at inference time remains identical to a standard item embedding lookup, while the information content approaches that of the original sample.




\subsection{Unified Architectures}

A recent generation of models unifies sequence modeling and feature interaction within a single Transformer backbone. \textbf{InterFormer}~\cite{interformer2023} interleaves FI and sequence layers. \textbf{MTGR}~\cite{mtgr2023} uses multi-task learning to jointly optimize sequence and FI objectives. \textbf{OneTrans}~\cite{onetrans2023} processes all feature fields as tokens in a single Transformer. \textbf{HyFormer}~\cite{hyformer2024}, our primary baseline, introduces a hybrid attention mechanism that separates intra-field and cross-field attention for efficiency. \textbf{MixFormer}~\cite{mixformer} combines MLP-Mixer and Transformer layers. The common limitation across all these architectures is that historical tokens remain at the item level.

\subsection{Vector Quantization for Recommendation}

VQ has seen growing adoption in recommender systems. For item tokenization, \textbf{VQ-Rec}~\cite{vqrec2023} learns discrete item codes for transferable recommendation, and \textbf{TIGER}~\cite{tiger2023} uses residual quantization for generative retrieval. In ranking, \textbf{UIST}~\cite{uist2024} pioneers discrete semantic tokenization for CTR prediction, \textbf{STORE}~\cite{store2025} introduces orthogonal rotation for semantic tokenization, \textbf{TRM}~\cite{trm2026} replaces item ID embeddings with semantic tokens at scale, \textbf{Zenith}~\cite{zenith2026} proposes Token Fusion and Token Boost for high-dimensional prime tokens, and \textbf{IAT}~\cite{iat2026} compresses full interaction features into unified instance embeddings. All these works quantize \emph{item} representations. SIF instead quantizes \emph{Raw Samples}---complete multi-field feature snapshots of historical interactions---requiring group-wise decomposition for heterogeneous fields and label-aware supervision, neither of which arises in item-VQ settings.

\begin{figure*}[t]
  \centering
  \includegraphics[width=0.92\textwidth]{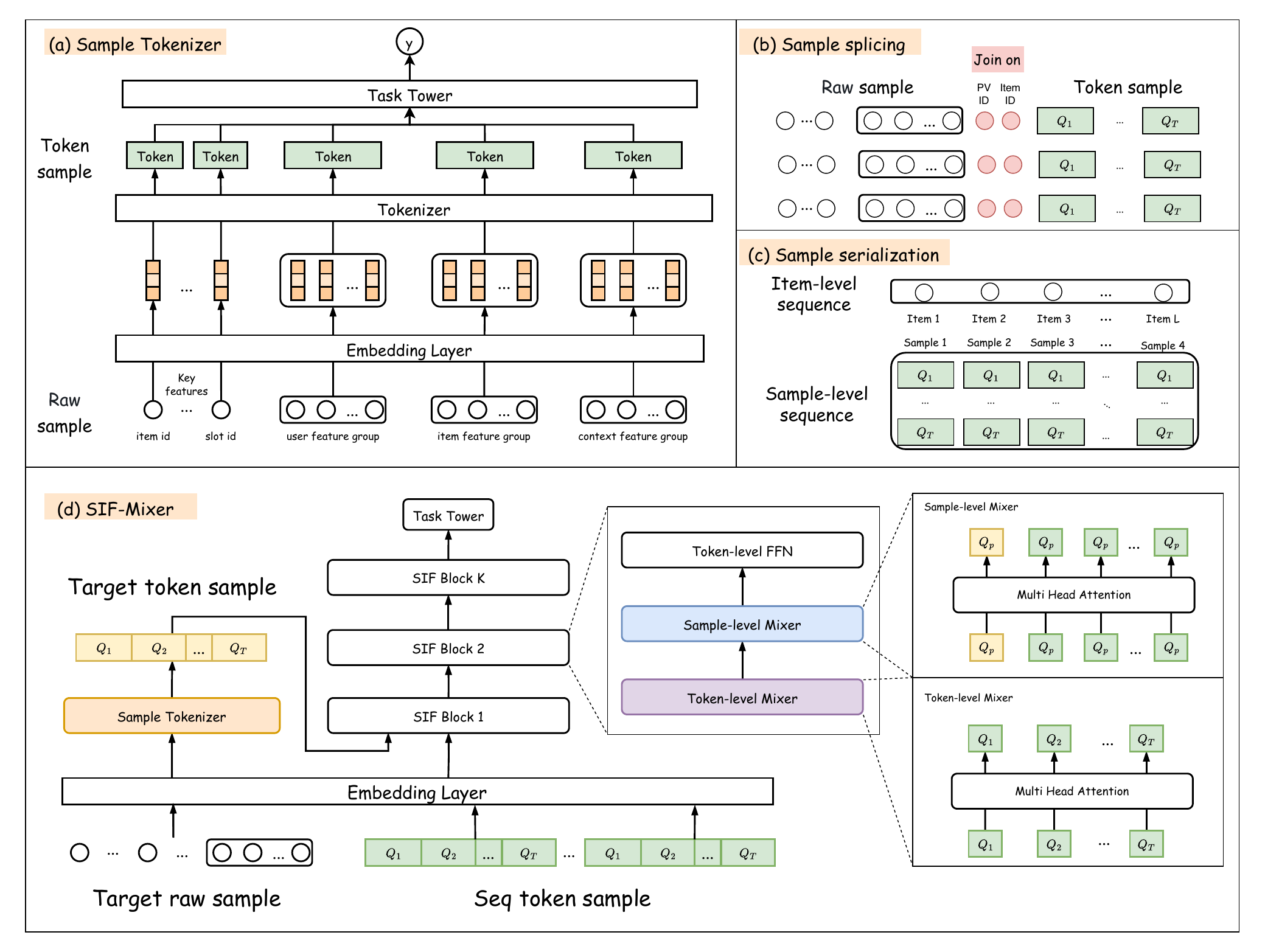}
  \caption{SIF overview. (a) Sample Tokenizer: HGAQ compresses Raw Sample $\mathcal{S}\to\mathcal{Q}$. (b) Sample splicing: item embeddings replaced by Token Samples. (c) Serialization with temporal position embeddings. (d) SIF-Mixer: $N$ blocks of Token-level Mixer, Sample-level Mixer, and FFN.}
  \label{fig:architecture}
\end{figure*}

\section{The SIF Framework}

\subsection{Sample-Level Token Definition}

\textbf{Core idea.} We define a \emph{Raw Sample} at interaction time $t$ as the complete feature tuple that characterized that historical event:
\begin{equation}
\mathcal{S} = [\mathbf{f}^{\text{user}} \mid \mathbf{f}^{\text{item}} \mid \mathbf{f}^{\text{ctx}} \mid \mathbf{f}^{\text{cross}}]
\end{equation}
where $\mathbf{f}^{\text{user}}$ are user profile features, $\mathbf{f}^{\text{item}}$ are item features, $\mathbf{f}^{\text{ctx}}$ are contextual signals, and $\mathbf{f}^{\text{cross}}$ are pre-computed cross features capturing joint user-item-context interaction patterns (e.g., user-category affinity, item-context co-occurrence statistics). The Raw Sample $\mathcal{S} \in \mathbb{R}^{d_s}$ is high-dimensional. The Sample Tokenizer compresses $\mathcal{S}$ into a Token Sample.

We distinguish two types of features in a sample: (1) \emph{point features}---user attributes, item attributes, and contextual signals---which are amenable to quantization; and (2) \emph{sequential features}---the user's historical interaction sequence $\mathcal{B}$---which carry temporal ordering information and are therefore excluded from sample quantization. Sequential features are modeled directly by the Sample-level Mixer in SIF-Mixer (\S4.6), which captures inter-sample dynamics via cross-temporal attention.

\subsection{Sample Tokenizer: Hierarchical Group-Adaptive Quantization}

The Sample Tokenizer $\mathcal{T}: \mathcal{S} \mapsto \mathcal{Q}$ maps each Raw Sample to a Token Sample $\mathcal{Q} = (q^{(g,k,m)})$---a compact set of $T \times M$ discrete codebook indices. At serving time only these indices need to be stored and retrieved; the corresponding embedding vectors are looked up on-the-fly from the codebook maintained in the SIF-Mixer.

\subsubsection{Group-Wise Decomposition}

Different components of $\mathcal{S}$ capture qualitatively different information. Rather than quantizing the entire snapshot with a single codebook, we first partition $\mathcal{S}$ into $G = 4$ semantic groups:
\begin{equation}
\mathcal{S} = [\underbrace{\mathbf{f}^{\text{user}}}_{G_1} \mid \underbrace{\mathbf{f}^{\text{item}}}_{G_2} \mid \underbrace{\mathbf{f}^{\text{ctx}}}_{G_3} \mid \underbrace{\mathbf{f}^{\text{cross}}}_{G_4}]
\end{equation}
This grouping is flexible: high-cardinality key features (e.g., item ID) that carry strong identity signals can be assigned to a dedicated singleton group, ensuring their discriminative information is not diluted by lower-cardinality features within a shared codebook.

\textbf{Adaptive intra-group sub-tokenization.} Within each semantic group $g$, the number of features $|\mathcal{F}_g|$ varies considerably. Compressing a large group into a single token forces semantically distinct sub-concepts into one vector, limiting the Token-level Mixer's ability to disentangle them. We therefore further divide each group into $K_g$ sub-tokens, where the number of sub-tokens is determined adaptively by the group's feature count:
\begin{equation}
  K_g = \left\lceil \frac{|\mathcal{F}_g|}{B} \right\rceil
  \label{eq:kg}
\end{equation}
where $B$ is the \emph{sub-token granularity}---the target number of feature fields assigned to each sub-token (default $B = 32$). Groups with more features naturally produce more sub-tokens, granting them proportionally more positions in the Token-level Mixer. The $|\mathcal{F}_g|$ features in group $g$ are evenly partitioned into $K_g$ non-overlapping subsets $\{\mathcal{F}_g^{(k)}\}_{k=1}^{K_g}$; each subset is projected to a fixed sub-token dimension $d_0$ via a group-and-slot-specific linear layer:
\begin{equation}
  \tilde{\mathbf{f}}^{(g,k)} = W^{(g,k)}_{\mathrm{proj}}\, \mathbf{f}^{(g,k)} \in \mathbb{R}^{d_0}
\end{equation}
where $\mathbf{f}^{(g,k)}$ denotes the raw features in the $k$-th subset of group $g$. All sub-tokens share the same dimension $d_0$ (default $d_0 = 16$), making the total token count per sample $T = \sum_{g=1}^{G} K_g$ and the total token embedding dimension $d = T \cdot d_0$. Each sub-token has its own codebook $\mathcal{C}_{g,k} = \{\mathbf{c}^{(g,k,m)}_v\}_{m,v}$ with $V = 256$ codes per level of dimension $d_0$.

\subsubsection{Residual Quantization}

Within each sub-token $(g, k)$, we apply $M$-level residual vector quantization (RVQ) to the projected sub-token vector $\tilde{\mathbf{f}}^{(g,k)}$:
\begin{equation}
q^{(g,k,m)} = \arg\min_v \|\mathbf{r}^{(g,k,m)} - \mathbf{c}^{(g,k,m)}_{v}\|_2, \quad \mathbf{r}^{(g,k,m)} = \mathbf{r}^{(g,k,m-1)} - \mathbf{c}^{(g,k,m-1)}_{q^{(g,k,m-1)}}
\end{equation}
where the initial residual $\mathbf{r}^{(g,k,1)} = \tilde{\mathbf{f}}^{(g,k)}$. The Token Sample concatenates all sub-token index sequences across all groups:
\begin{equation}
\mathcal{Q} = \bigl[(q^{(g,k,1)},\ldots,q^{(g,k,M)})\bigr]_{g=1,\ldots,G;\; k=1,\ldots,K_g}
\end{equation}
With $G=4$, $B=32$, $M=3$, $V=256$: the total number of sub-tokens is $T = \sum_g K_g$ (e.g.\ $T{=}27$ under our feature schema), and each $\mathcal{S}$ is compressed to $T \times M \times \log_2 V = 27{\times}3{\times}8 = 648$ bits.

\subsubsection{Label-Supervised Codebook Training}

The Sample Tokenizer is jointly trained with the ranking objective. During training, the RVQ reconstruction $\hat{\mathbf{s}}$ is fed to a lightweight MLP to predict CTR:
\begin{equation}
\hat{\mathbf{s}} = \Bigl[\textstyle\sum_m \mathbf{c}^{(g,k,m)}_{q^{(g,k,m)}}\Bigr]_{g,k},\qquad \hat{y}^{\text{token}} = \sigma\bigl(\text{MLP}(\hat{\mathbf{s}})\bigr)
\end{equation}
This auxiliary pCTR loss $\mathcal{L}_{\text{token}} = \mathcal{L}_{\text{BCE}}(\hat{y}^{\text{token}}, y)$ is added to the overall training objective, ensuring that the codebook is organized by predictive relevance rather than purely by reconstruction error. Importantly, for important categorical features such as item IDs, we retain a dedicated per-field quantization group to preserve individual item identity, preventing high-cardinality identifiers from being merged with less-discriminative features within a shared group.



\subsection{Sample Splicing and Serialization}
\label{sec:splicing}

Standard behavior sequences record only item IDs. SIF upgrades each entry with the full feature snapshot and compresses it into a Token Sample, transforming the sequence from item-level to sample-level:
\begin{equation}
\underbrace{\{i_1,\;\ldots,\;i_L\}}_{\text{item sequence (standard)}}
\;\xrightarrow{\;\mathcal{T}\;}
\underbrace{\{\mathcal{Q}_1,\;\ldots,\;\mathcal{Q}_L\}}_{\text{sample sequence (SIF)}}
\end{equation}
Each $\mathcal{Q}_l \in \{1,\ldots,V\}^{T\times M}$ encodes the complete multi-field snapshot of the $l$-th interaction as $T{\times}M$ discrete indices ($T{\times}M{\times}8$ bits at $V{=}256$), a substantially more compact representation than the original float32 features. The Token Samples are pre-computed offline and stored in a key-value store; the sample sequence, ordered chronologically, constitutes the offline-prepared input to SIF-Mixer.

\subsection{SIF-Mixer}

\textbf{SIF-Mixer} is a backbone architecture, inspired by MLP-Mixer-style factored designs~\cite{mixformer}, that stacks $N$ identical \textbf{SIF Blocks}. Each block decomposes the interaction into three sequential sub-operations: Token-level Mixer, Sample-level Mixer, and Token-level FFN. This factored design explicitly separates intra-sample feature interaction from inter-sample temporal interaction, reducing attention complexity while preserving both interaction types.

\textbf{Input layout.} The input to SIF-Mixer consists of two parts, which together form the initial hidden state $\mathbf{H}^0 \in \mathbb{R}^{(L+1)\times T \times d_0}$, where $T = \sum_{g=1}^{G} K_g$ is the total number of sub-tokens per sample.

\textit{Seq Token Samples} ($l=1,\ldots,L$): the sample sequence $\{\mathcal{Q}_1,\ldots,\mathcal{Q}_L\}$ from \S\ref{sec:splicing} is embedded via codebook lookup with recency encoding. For the sub-token at position $(g,k)$:
\begin{equation}
\mathbf{z}^{(g,k)}_l = \sum_{m=1}^{M} \mathbf{c}^{(g,k,m)}_{q^{(g,k,m)}_l} \in\mathbb{R}^{d_0}, \qquad
\mathbf{H}^0_{l,*} = \bigl[\mathbf{z}^{(1,1)}_l \;\|\; \cdots \;\|\; \mathbf{z}^{(G,K_G)}_l\bigr] + \mathbf{p}_{L-l}
\end{equation}

\textit{Target Token Sample} ($l=0$): unlike historical Token Samples, the current-request features $\{\mathbf{f}^{(g)}_{\tau}\}$ are only available at serving time and cannot be pre-quantized offline. They are instead projected per sub-token into the same codebook space on-the-fly via a learned linear projection:
\begin{equation}
\mathbf{H}^0_{0,(g,k)} = W_{\mathrm{res}}^{(g,k)}\mathbf{f}^{(g,k)}_{\tau}
\end{equation}
$\mathcal{L}_{\text{align}}$ ensures each $W_{\mathrm{res}}^{(g,k)}$ maps into the same space as its codebook, making all $L+1$ rows compatible for cross-temporal attention.

\textbf{SIF Block.} Each block $n \in \{1,\ldots,N\}$ operates on $\mathbf{H}^{n-1}$ viewed as a $(L+1)\times T$ matrix of $d_0$-dimensional vectors---rows are samples, columns are sub-token positions ($T$ sub-tokens from all groups):
\begin{equation}
\mathbf{H}^n \;=\;
\overbrace{
\left[\begin{array}{ccc}
\mathbf{H}^n_{0,1} & \cdots & \mathbf{H}^n_{0,T} \\[2pt]
\mathbf{H}^n_{1,1} & \cdots & \mathbf{H}^n_{1,T} \\[2pt]
\vdots                & \ddots & \vdots                \\[2pt]
\mathbf{H}^n_{L,1} & \cdots & \mathbf{H}^n_{L,T}
\end{array}\right]
}^{\displaystyle\longleftarrow\; T \text{ sub-tokens}\;\longrightarrow\;\textit{Token-level Mixer}}
\left.\vphantom{\begin{array}{c}\\[2pt]\\[2pt]\\[2pt]\\[2pt]\end{array}}\right\}
\!\!\begin{array}{l}\scriptstyle L{+}1\\\scriptstyle\text{samples}\\\scriptstyle\textit{Sample-}\\\scriptstyle\textit{level}\\\scriptstyle\textit{Mixer}\end{array}
\end{equation}
Three sub-operations are applied in sequence (MHA: multi-head attention~\cite{attention2017}; LN: pre-norm LayerNorm~\cite{layernorm2016}):

\noindent\textit{(i) Token-level Mixer} runs self-attention along each \emph{row}, modeling interactions among all $T$ sub-tokens within a single sample. Because sub-tokens from different groups (and different sub-positions within the same group) occupy distinct column positions, the attention can capture both inter-group correlations (e.g.\ user--item--context) and intra-group sub-concept interactions (e.g.\ price signals interacting with review signals within the item group):
\begin{equation}
\tilde{\mathbf{H}}^n_l = \mathbf{H}^{n-1}_l + \text{MHA}\bigl(\text{LN}(\mathbf{H}^{n-1}_l)\bigr), \quad l = 0,\ldots,L
\end{equation}

\noindent\textit{(ii) Sample-level Mixer} runs self-attention along each \emph{column}, modeling interactions among all $L+1$ samples at the same sub-token position---crucially, the Target Token Sample ($l=0$) can attend to all Seq Token Samples to extract relevant historical context:
\begin{equation}
\bar{\mathbf{H}}^n_{*,p} = \tilde{\mathbf{H}}^n_{*,p} + \text{MHA}\bigl(\text{LN}(\tilde{\mathbf{H}}^n_{*,p})\bigr), \quad p = 1,\ldots,T
\end{equation}

\noindent\textit{(iii) Token-level FFN} applies a position-wise non-linear transformation to each entry:
\begin{equation}
\mathbf{H}^n_{l,p} = \bar{\mathbf{H}}^n_{l,p} + \text{FFN}\bigl(\text{LN}(\bar{\mathbf{H}}^n_{l,p})\bigr)
\end{equation}

\textbf{Prediction head.} After $N$ SIF Blocks, we extract the target sample representation by mean-pooling its $T$ sub-token outputs:
\begin{equation}
\mathbf{h} = \frac{1}{T}\sum_{p=1}^{T} \mathbf{H}^N_{0,p}
\end{equation}
The ranking score is produced by a two-layer MLP with sigmoid output:
\begin{equation}
\hat{y} = \sigma\!\left(\mathbf{w}_2^\top \text{ReLU}(\mathbf{W}_1 \mathbf{h} + \mathbf{b}_1) + b_2\right)
\end{equation}

\textbf{Training objective.}\label{sec:training} The overall loss is:
\begin{equation}
\mathcal{L} = \mathcal{L}_{\text{BCE}} + \beta\,\mathcal{L}_{\text{VQ}} + \gamma\,\mathcal{L}_{\text{align}}
\end{equation}
where $\beta = 1.0$ and $\gamma = 0.25$. $\mathcal{L}_{\text{VQ}}$ is the standard VQ commitment loss~\cite{vqvae2017} ($\lambda{=}0.25$) between the group encoder output and the RVQ reconstruction. The alignment loss trains the target token projection $W_{\mathrm{res}}^{(g)}$ to be consistent with the codebook space, so that target and historical token representations are compatible at inference time:
\begin{equation}
\label{eq:align}
\mathcal{L}_{\text{align}} = \sum_{g=1}^{G}\sum_{k=1}^{K_g} \left\| W_{\mathrm{res}}^{(g,k)}\mathbf{f}^{(g,k)}_{\tau} - \mathrm{sg}(\mathbf{e}_{g,k}) \right\|^2
\end{equation}
where $\mathrm{sg}(\cdot)$ denotes stop-gradient and $\mathbf{e}_{g,k} = \sum_m \mathbf{c}^{(g,k,m)}_{q^{(g,k,m)}}$ is the Tokenizer's codebook reconstruction for sub-token $(g,k)$. During training, $\mathbf{e}_{g,k}$ is obtained by passing the current-request Raw Sample $\mathcal{S}_\tau$ through the Sample Tokenizer to produce the RVQ reconstruction $\mathbf{e}_{g,k} = \sum_m \mathbf{c}^{(g,k,m)}_{q^{(g,k,m)}_\tau}$, which serves as the alignment target; at serving time this Tokenizer forward pass is skipped and only $W_{\mathrm{res}}^{(g,k)}$ is applied. Because each Raw Sample $\mathcal{S}$ incorporates the behavioral outcome of the original interaction, the ranking supervision signal $\mathcal{L}_{\text{BCE}}$ propagates through VQ into codebook learning, organizing snapshots by their predictive context without requiring an additional contrastive term. \textbf{Complexity.} Each SIF Block incurs $O(T^2 \cdot (L+1) \cdot d_0 + (L+1)^2 \cdot T \cdot d_0)$ attention operations. Since $T \ll L+1$ in practice (a modest number of sub-tokens vs.\ hundreds of history steps), the sample-level mixer dominates: $O(L^2 \cdot T \cdot d_0)$, matching the complexity of standard sequence attention at the same sequence length. The adaptive sub-tokenization increases $T$ relative to the fixed-$G$ design, but $T$ remains small (typically $T < 20$ under $B{=}8$), keeping the token-level mixer overhead negligible.

\section{Experiments}

\subsection{Experimental Setup}

\textbf{Datasets.} We evaluate SIF on one large-scale industrial dataset:
\textbf{Industrial Dataset}: 1B+ impression records (spanning 90 days), 50M+ users, 5M+ items. Each sample includes ${\sim}$600+ feature fields spanning user profile, item attributes, contextual signals, and a 1000-item behavior sequence. All 4 semantic groups ($G_1$--$G_4$) with adaptive sub-tokenization ($B{=}32$) are fully utilized.

All baseline models receive the same current-request feature set as SIF (user fields, item fields, and contextual fields for the target request); the sole difference across methods is the representation of the \emph{historical behavior sequence}---baselines use standard item embeddings, while SIF replaces each historical token with an HGAQ-compressed Token Sample.

\textbf{Evaluation Metrics.} We report \textbf{AUC}, \textbf{GAUC} (Group AUC), and \textbf{FLOPs} on the industrial dataset. All results are averaged over 5 independent runs; statistical significance by paired $t$-test ($p < 0.01$).

\textbf{Baselines.} Each model is characterized by its sequence modeling component and feature interaction component. We organize comparisons along two axes:
\begin{itemize}
  \item \emph{Varying feature interaction}: \textbf{DCNv2}~\cite{dcnv22021} (base), \textbf{Wukong}~\cite{wukong2023}, \textbf{RankMixer}~\cite{rankmixer}.
  \item \emph{Varying sequence modeling}: \textbf{DIN}~\cite{din2018}, \textbf{SIM}~\cite{sim2020}, \textbf{LONGER}~\cite{longer}.
  \item \emph{Unified frameworks (both components jointly optimized)}: \textbf{HyFormer}~\cite{hyformer2024}, \textbf{OneTrans}~\cite{onetrans2023}.
\end{itemize}

\textbf{Implementation Details.} All models are implemented in PyTorch on 8$\times$A100-80G GPUs. For the SIF-Mixer: $N = 4$ SIF Blocks, 8 attention heads, sub-token dimension $d_0 = 16$, FFN dimension $4{\times}d_0$, pre-norm LayerNorm, mean pooling over $T$ sub-tokens. The Sample Tokenizer: $G = 4$, $B = 32$, $M = 3$, $V = 256$, $d_0 = 16$. Optimizer: Adam~\cite{adam2015} (lr=$10^{-3}$, $\beta_1=0.9$, $\beta_2=0.999$, weight decay $10^{-5}$), batch size 4096. Sequence length $L = 1000$.

\subsection{Main Results}

Table~\ref{tab:main} presents the main comparison results. Bold indicates the best result; $\ddagger$ denotes statistically significant improvement over HyFormer ($p < 0.01$, paired $t$-test, 5 runs).

\begin{table*}[t]
  \caption{Offline results on the industrial dataset. Higher AUC/GAUC is better.}
  \label{tab:main}
  \centering
  \resizebox{\textwidth}{!}{%
  \begin{tabular}{ll|cc|cc|cc}
    \toprule
    \multirow{2}{*}{\textbf{Feature Interaction}} & \multirow{2}{*}{\textbf{Sequence Modeling}}
      & \multicolumn{2}{c|}{\textbf{CTR}}
      & \multicolumn{2}{c|}{\textbf{CVR (order)}}
      & \multicolumn{2}{c}{\textbf{Efficiency}} \\
    \cmidrule(lr){3-4}\cmidrule(lr){5-6}\cmidrule(lr){7-8}
     & & AUC$\uparrow$ & GAUC$\uparrow$ & AUC$\uparrow$ & GAUC$\uparrow$ & Params (M) & TFLOPs \\
    \midrule
    DCNv2~\cite{dcnv22021} & DIN~\cite{din2018}
      & 0.7832 & 0.7614 & 0.8103 & 0.7891 & 48 & 0.31 \\
    \midrule
    Wukong~\cite{wukong2023} & SIM~\cite{sim2020}
      & +0.41\% & +0.38\% & +0.35\% & +0.33\% & 56 & 0.38 \\
    Wukong~\cite{wukong2023} & LONGER~\cite{longer}
      & +0.53\% & +0.49\% & +0.44\% & +0.41\% & 62 & 0.42 \\
    RankMixer~\cite{rankmixer} & SIM~\cite{sim2020}
      & +0.67\% & +0.61\% & +0.58\% & +0.54\% & 51 & 0.35 \\
    RankMixer~\cite{rankmixer} & LONGER~\cite{longer}
      & +0.79\% & +0.72\% & +0.68\% & +0.63\% & 53 & 0.40 \\
    \midrule
    \multicolumn{2}{l|}{HyFormer~\cite{hyformer2024} }
      & +1.12\% & +1.01\% & +0.97\% & +0.88\% & 120 & 0.87 \\
    \multicolumn{2}{l|}{OneTrans~\cite{onetrans2023} }
      & +1.08\% & +0.96\% & +0.91\% & +0.83\% & 115 & 0.82 \\
    \midrule
    \multicolumn{2}{l|}{\textbf{SIF (Ours)}}
      & \textbf{+2.03\%} & \textbf{+1.89\%} & \textbf{+1.74\%} & \textbf{+1.61\%} & \textbf{128} & \textbf{0.93} \\
    \bottomrule
  \end{tabular}}%
\end{table*}

\textbf{Analysis.} SIF outperforms all baselines across all metrics, with improvements that are statistically significant ($p < 0.01$) in every setting.

\textbf{(1) SIF vs.\ unified baselines.} SIF outperforms HyFormer by +0.91\% AUC and +0.88\% GAUC on CTR, and by +0.77\% AUC and +0.73\% GAUC on CVR ($p < 0.01$). Elevating sequence tokens from item-level to sample-level delivers a substantial and reliable gain beyond the unified Transformer architecture itself.

\textbf{(2) Specialized models vs.\ unified frameworks.} Among specialized models, sequence-modeling approaches (SIM, LONGER) consistently outperform feature-interaction approaches (Wukong, RankMixer) on CTR, confirming the importance of sequential modeling in behavior-rich industrial settings. Unified frameworks (HyFormer, OneTrans) further outperform all specialized models by jointly modeling sequence dynamics and feature interactions, validating the architectural paradigm of unification.

\textbf{(3) Industrial significance.} The observed AUC improvement of SIF over HyFormer (+0.91\% relative, +0.0071 absolute on CTR AUC) is practically meaningful at industrial scale. A widely validated empirical rule-of-thumb~\cite{din2018,twin2023,wukong2023} is that 0.001 absolute AUC corresponds to approximately 0.1\%+ CTR improvement in online A/B experiments, projecting SIF's gains to +0.7\%+ CTR improvement. The observed online gain (+2.03\% CTR, \S\ref{sec:online}) substantially exceeds this projection, suggesting that sample-level token enrichment provides complementary signals---such as time-varying item popularity and contextual demand shifts---that are not captured by static AUC evaluation.

\subsection{Ablation Study}

We organize the ablation into two parts: (1) ablation of the Sample Tokenizer, evaluating different token representations along both quality (GAUC) and compression efficiency (compression ratio); and (2) ablation of the SIF-Mixer backbone, isolating the contribution of each mixing component.

\subsubsection{Sample Tokenizer Ablation}
\label{sec:tokenizer-ablation}

Table~\ref{tab:ablation} compares four token representation strategies. Metrics are the GAUC gap relative to SIF full on CTR and CVR, and the \emph{storage compression ratio}, defined as the ratio of the original raw snapshot storage to the per-sample token storage:
\begin{equation}
  \text{Storage Compression Ratio} = \frac{b_{\text{snapshot}}}{b_{\text{token}}},
  \label{eq:compression}
\end{equation}
where $b_{\text{snapshot}}$ is the bit-size of the full non-item snapshot stored in float32 ($|\mathcal{F}_{\text{non-seq}}| \times d_e \times 32$~bits), and $b_{\text{token}}$ is the bit-size of the stored token. In our setup, $|\mathcal{F}_{\text{non-seq}}|{=}600$, $d_e{=}8$, giving $b_{\text{snapshot}}{=}153{,}600$~bits. For HGAQ, only $T{\times}M$ discrete indices (each $\log_2 V = 8$~bits) are stored: $b_{\text{token}} = 27{\times}3{\times}8 = 648$~bits, yielding a storage compression ratio of ${\approx}237{\times}$. Crucially, this high ratio is achieved through genuine compression---the full snapshot context is recovered at inference via codebook lookup---rather than by discarding features.

\begin{table}[t]
  \caption{Sample Tokenizer ablation. $\Delta$GAUC relative to SIF; compression ratio = raw bits / token bits.}
  \label{tab:ablation}
  \centering
  \small
  \resizebox{\columnwidth}{!}{%
  \begin{tabular}{lccc}
    \toprule
    \textbf{Token Representation} & $\Delta$\textbf{CTR-GAUC} & $\Delta$\textbf{CVR-GAUC} & \textbf{Comp.~Ratio} \\
    \midrule
    \textbf{SIF (HGAQ token, ours)}       & ---         & ---         & $\approx$237$\times$ \\
    Item ID only                          & $-$1.00\%  & $-$0.86\%  & $\approx$2400$\times$ \\
    Item ID + key features                & $-$0.60\%  & $-$0.51\%  & $\approx$185$\times$ \\
    Raw sample emb ($d{=}512$, dense)     & $-$0.27\%  & $-$0.23\%  & $\approx$9$\times$ \\
    \bottomrule
  \end{tabular}}%
\end{table}

\textit{Variant descriptions.}
\begin{description}[leftmargin=0pt,labelindent=0pt,itemsep=1pt,parsep=0pt]
  \item[\emph{Item ID only}:] each historical position is represented by a single item-ID embedding, equivalent to a standard item-embedding sequence backbone. Cross-feature groups and the Token-level Mixer are inapplicable and thus disabled. Storage: 1 int64 item ID index $= 64$~bits ($\approx$2400$\times$ vs.~raw snapshot).
  \item[\emph{Item ID + key features}:] item-ID embedding concatenated with a small set of high-importance handcrafted features (e.g., price bucket, category, CTR statistics), projected to the same token dimension. Storage: 1 int64 item ID index + $\sim$24 int32/float32 scalar fields $= 64 + 24{\times}32 = 832$~bits ($\approx$185$\times$ vs.~raw snapshot).
  \item[\emph{Raw sample emb ($d{=}512$, dense)}:] all raw snapshot features concatenated and projected via a linear layer to a 512-dim dense vector, without any quantization. Storage: 512 float32 values $= 512{\times}32 = 16{,}384$~bits ($\approx$9$\times$ vs.~raw snapshot). Token-level Mixer applies but quantization structure is absent.
  \item[\emph{SIF (HGAQ token, ours)}:] only $T{\times}M$ discrete codebook indices are stored offline; at serving time embeddings are retrieved via codebook lookup. Storage: $27{\times}3{\times}8 = 648$~bits ($\approx$237$\times$ vs.~raw snapshot), with full snapshot semantics recovered on-the-fly.
\end{description}

\textbf{Analysis.} Item ID only performs worst ($-$1.00\% CTR-GAUC, $-$0.86\% CVR-GAUC), falling below OneTrans (Table~\ref{tab:main}), because the entire non-item snapshot context is discarded and the Token-level Mixer cannot be applied. Although its storage ratio ($\approx$2400$\times$) appears high, this reflects information loss rather than genuine compression---the entire non-item context is simply discarded. Adding key features recovers a portion of the gap ($-$0.60\%/$-$0.51\%) at $\approx$185$\times$, but still discards the majority of contextual signals. Notably, HGAQ achieves a higher storage ratio ($\approx$237$\times$) than ``Item ID + key features'' while retaining the full snapshot context of 800+ feature fields.

The dense raw-sample embedding ($d{=}512$) retains all snapshot features yet still underperforms HGAQ ($-$0.27\%/$-$0.23\%), despite having a comparable storage footprint ($\approx$9$\times$ compression). Higher information volume does not directly translate to better sequence modeling. Three factors explain the quality gap. First, a 512-dim token is comparable in size to an HGAQ token ($T{\times}d_0 = 432$ dim), but lacks discrete structure, making cross-temporal attention over $L{=}1000$ positions significantly harder to optimize---larger token dimensions enlarge the parameter space, increase gradient variance, and slow convergence. Second, HGAQ's discrete codebook imposes an implicit clustering constraint: historically similar snapshots map to identical or nearby codes, providing a natural regularization that an unconstrained linear projection lacks. Third, because all historical positions share the same codebook, HGAQ tokens are semantically aligned across time, enabling the Token-level and Sample-level Mixers to learn cross-temporal patterns more effectively. In short, HGAQ trades raw information volume for \emph{learnability}: the structured, compact, and temporally aligned token representation proves more useful to the downstream Mixer than a higher-dimensional but unstructured dense vector.

\begin{figure*}[t]
  \centering
  \subfigure[CTR GAUC vs.\ granularity $B$. Red dot: $B{=}32$.]{%
    \includegraphics[width=0.31\textwidth]{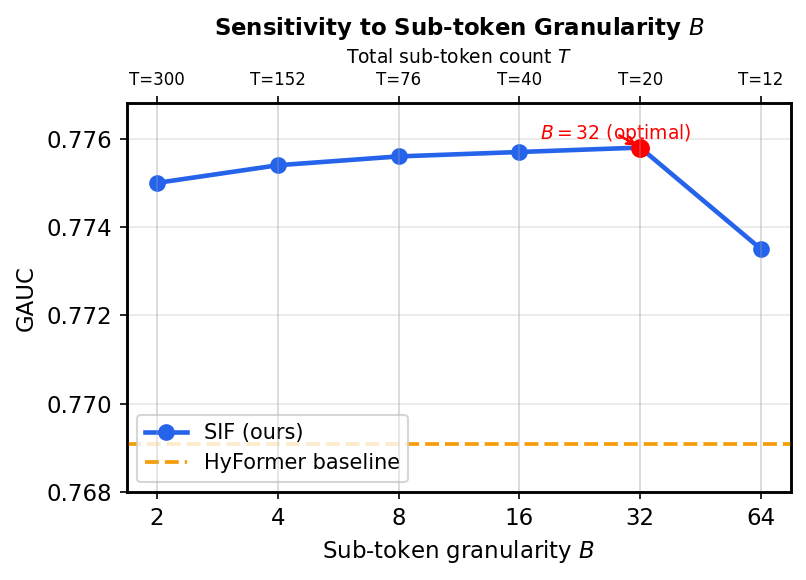}%
    \label{fig:k_sweep}%
  }\hfill
  \subfigure[CTR GAUC vs.\ TFLOPs (depth $N$). Red dot: $N{=}4$.]{%
    \includegraphics[width=0.31\textwidth]{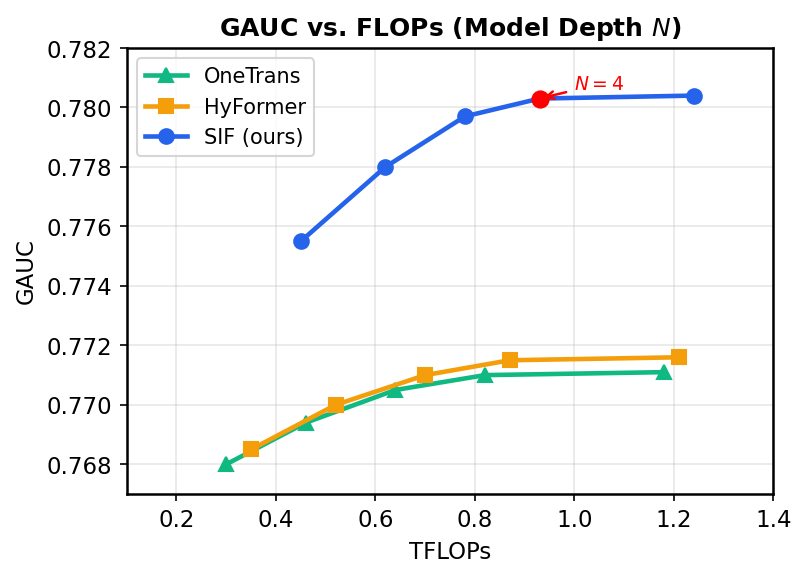}%
    \label{fig:flops_scaling}%
  }\hfill
  \subfigure[CTR GAUC vs.\ sequence length $L$.]{%
    \includegraphics[width=0.31\textwidth]{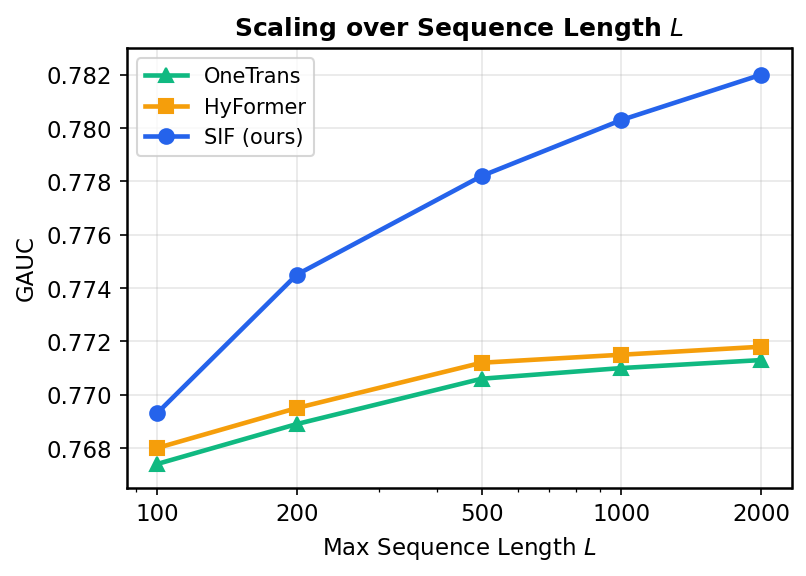}%
    \label{fig:scaling}%
  }
  \caption{Scaling analysis: (a) sub-token granularity $B$, (b) model depth $N$, (c) sequence length $L$.}
  \label{fig:analysis}
\end{figure*}

\textbf{Sensitivity to sub-token granularity $B$.} We sweep the sub-token granularity $B \in \{2, 4, 8, 16, 32, 64\}$, which controls how many sub-tokens each group produces via $K_g = \lceil |\mathcal{F}_g| / B \rceil$, giving $T \approx \lceil 600 / B \rceil$ total sub-tokens (Figure~\ref{fig:k_sweep}). $B = 32$ achieves the best GAUC (0.7758), striking the optimal balance between intra-group disentanglement and model compactness. At $B = 2$ (fine-grained sub-tokenization, $T{=}300$), GAUC drops slightly to 0.7750 due to excessive token sequence length making the Token-level Mixer harder to optimize. At $B = 64$ (coarse sub-tokenization, $T{=}12$), GAUC falls to 0.7735, approaching the behavior of the fixed-$G$ design with insufficient intra-group decomposition. SIF outperforms HyFormer ($\text{GAUC}{=}0.7691$) for all tested values of $B$, demonstrating robustness across a wide range of granularity choices. We adopt $B = 32$ ($T{=}20$) as the default.

\subsubsection{SIF-Mixer Architecture Ablation}
\label{sec:mixer-ablation}

Given that each sequence position carries $T$ side-information sub-tokens, there are multiple ways to apply attention over the resulting $(L{+}1){\times}T$ token matrix. Table~\ref{tab:mixer-ablation} compares three strategies on the industrial dataset (5 runs each) to identify which attention pattern best exploits the sample-level token structure. Metrics are the GAUC gap relative to SIF full on CTR and CVR.

\begin{table}[h]
  \caption{SIF-Mixer attention strategy ablation. $\Delta$GAUC relative to SIF full.}
  \label{tab:mixer-ablation}
  \centering
  \small
  \resizebox{\columnwidth}{!}{%
  \begin{tabular}{lcc}
    \toprule
    \textbf{Attention Strategy} & $\Delta$\textbf{CTR-GAUC} & $\Delta$\textbf{CVR-GAUC} \\
    \midrule
    \textbf{SIF-Mixer (factored row+col)}  & ---                       & --- \\
    Flat attention                         & $-$0.24\%$\pm$0.01\%   & $-$0.20\%$\pm$0.01\% \\
    Pooled-then-attend                     & $-$0.81\%$\pm$0.02\%   & $-$0.68\%$\pm$0.02\% \\
    \bottomrule
  \end{tabular}}%
\end{table}

\textit{Strategy descriptions.}
\begin{description}[leftmargin=0pt,labelindent=0pt,itemsep=1pt,parsep=0pt]
  \item[\emph{Flat attention}:] all $(L{+}1){\times}T$ sub-tokens are flattened into a single sequence and processed by standard full self-attention, with complexity $O((LT)^2)$.
  \item[\emph{Pooled-then-attend}:] each sample's $T$ sub-tokens are first mean-pooled into a single vector, then standard sequence attention is applied across $L{+}1$ pooled representations, with complexity $O(L^2)$. This is the typical approach in existing sequence models that incorporate side information.
  \item[\emph{SIF-Mixer (factored row+col)}:] intra-sample Token-level Mixer (row attention, $O(T^2 \cdot L)$) followed by inter-sample Sample-level Mixer (column attention, $O(L^2 \cdot T)$), total $O(L^2 T + LT^2) \approx O(L^2 T)$.
\end{description}

\textbf{Analysis.} Pooled-then-attend ($-$0.81\%/$-$0.68\%) suffers the largest drop: collapsing $T$ sub-tokens into a single vector before attention discards intra-sample feature structure, preventing the model from capturing fine-grained interactions among user, item, context, and cross features within each historical snapshot. Its absolute CTR-GAUC ($+$1.08\% over DCNv2+DIN) is only marginally above HyFormer ($+$1.01\%), confirming that the gain from richer Token Sample inputs is largely wasted when sub-token structure is discarded via pooling. Flat attention ($-$0.24\%/$-$0.20\%) recovers most of the gap by attending over all sub-tokens jointly, but lacks the explicit row/column inductive bias and incurs $O((LT)^2)$ complexity that becomes prohibitive at $L{=}1000$, $T{=}27$. The factored SIF-Mixer achieves the best quality at lower cost: by decomposing attention into intra-sample and inter-sample axes, it fully exploits the two-dimensional structure of the token matrix while keeping complexity linear in $T$.

\subsection{Scaling Analysis}

We study two dimensions of SIF's scaling behavior: model depth and sequence length.

\paragraph{Model Depth ($N$).} Figure~\ref{fig:flops_scaling} plots GAUC against TFLOPs as the number of layers $N$ is varied from 1 to 6 for all three models. SIF achieves a consistently better GAUC-FLOPs trade-off across the entire depth range: at matched FLOPs (0.87 TFLOPs, $N=4$), SIF reaches GAUC 0.7803 vs.\ HyFormer's 0.7715 (+0.0088) and OneTrans's 0.7710 (+0.0093), consistent with the default-configuration results in Table~\ref{tab:main}. HyFormer saturates early due to the heavy per-layer cost of full attention, while OneTrans plateaus at a lower GAUC ceiling. SIF continues to improve until $N=4$ and we adopt this as the default.

\paragraph{Sequence Length.}A key property of SIF is whether its advantage persists or grows as the history budget increases. We evaluate SIF, HyFormer, and OneTrans across $L \in \{100, 200, 500, 1000, 2000\}$ on the industrial dataset (default $L{=}1000$, consistent with Table~\ref{tab:main}).

\begin{table}[t]
  \caption{CTR GAUC vs.\ sequence length $L$ (mean over 5 runs).}
  \label{tab:scaling}
  \centering
  \small
  \resizebox{\columnwidth}{!}{%
  \begin{tabular}{lcccc}
    \toprule
    \textbf{Seq. Length $L$} & \textbf{OneTrans} & \textbf{HyFormer} & \textbf{SIF (ours)} & $\Delta$\textbf{(SIF$-$Hyf)} \\
    \midrule
    100  & 0.7674 & 0.7680 & 0.7693 & +0.0013 \\
    200  & 0.7689 & 0.7695 & 0.7745 & +0.0050 \\
    500  & 0.7706 & 0.7712 & 0.7782 & +0.0070 \\
    1000 & 0.7710 & 0.7715 & 0.7803 & +0.0088 \\
    2000 & 0.7713 & 0.7718 & 0.7820 & +0.0102 \\
    \bottomrule
  \end{tabular}}%
\end{table}

All three models improve with longer sequences; however, the gains diverge substantially. HyFormer and OneTrans saturate quickly: from $L=500$ to $L=2000$ the GAUC improvement is only +0.0005 and +0.0006 respectively, consistent with the encoding bottleneck in item-level sequence architectures. All three models benefit substantially from longer sequences, but SIF's gains are markedly steeper. At $L{=}100$, SIF (0.7693) is already competitive with HyFormer at $L{=}200$ (0.7695), and by $L{=}500$ SIF (0.7782) surpasses HyFormer at $L{=}1000$ (0.7715). The performance gap over HyFormer widens monotonically from +0.0013 at $L=100$ to +0.0102 at $L=2000$. This reflects SIF's token design: each additional history position contributes a fully-contextualized Raw Sample to cross-temporal attention, while item-level methods compress each position to a bare item embedding and hit a representational ceiling. Figure~\ref{fig:scaling} visualizes these trends.






\subsection{Online A/B Experiment}
\label{sec:online}

SIF is deployed in an industrial local-service recommendation pipeline. Results from a \textbf{5\% traffic holdout over 7 days} show improvements of \textbf{+2.03\% CTR}, \textbf{+1.21\% CVR}, and \textbf{+1.35\% GMV/session} compared to the HyFormer production baseline.

To understand how SIF's benefit depends on users' history richness, we further stratify the online result by \textbf{behavior sequence length} $L$---the number of logged interactions available for each user. Table~\ref{tab:online_seqlen} reports the per-stratum CTR, CVR, and GMV/session lifts.

\begin{table}[h]
  \caption{Online A/B results by sequence length $L$ (5\% holdout, 7 days) vs.\ HyFormer baseline.}
  \label{tab:online_seqlen}
  \centering
  \small
  \resizebox{\columnwidth}{!}{%
  \begin{tabular}{lccc}
    \toprule
    \textbf{Sequence Length} & $\Delta$\textbf{CTR} & $\Delta$\textbf{CVR} & $\Delta$\textbf{GMV/session} \\
    \midrule
    $L < 10$   (cold users)  & +0.53\% & +0.31\% & +0.37\% \\
    $10  \le L < 100$         & +1.18\% & +0.71\% & +0.84\% \\
    $100 \le L < 500$         & +2.07\% & +1.24\% & +1.38\% \\
    $L \ge 500$ (heavy users) & +3.12\% & +1.87\% & +2.06\% \\
    \midrule
    \textbf{Overall}           & \textbf{+2.03\%} & \textbf{+1.21\%} & \textbf{+1.35\%} \\
    \bottomrule
  \end{tabular}}%
\end{table}

The gains scale monotonically with $L$ across all three metrics. Heavy users ($L \ge 500$) benefit most (+3.12\% CTR, +1.87\% CVR, +2.06\% GMV/session), as the Sample-level Mixer can leverage a richer pool of fully-contextualized Token Samples for cross-temporal reasoning. Even cold users ($L < 10$) see a meaningful improvement (+0.53\% CTR, +0.31\% CVR), which we attribute primarily to the \textbf{Sample Tokenizer} rather than the sequence modeling component: by encoding the current-request features into the same codebook space as historical Token Samples, the Target Token Sample representation becomes more expressive and semantically aligned even without a long history---a benefit that is orthogonal to sequence length. The improvement is consistent across all user segments, with the largest gains for users with $L \geq 500$.

\section{Conclusion}

We presented \textbf{SIF} (\emph{Sample Is Feature}), a framework that elevates each historical sequence token from a bare item embedding to a full sample-level representation. Every past interaction already has a complete request record in the training log---yet existing architectures discard it in favor of a bare item ID. SIF closes this gap via the \textbf{Sample Tokenizer} (offline HGAQ compression of Raw Samples into compact Token Samples) and the \textbf{SIF-Mixer} (factored token-level and sample-level mixing over homogeneous representations). Offline experiments show up to \textbf{+0.88\% GAUC} over the best unified baseline; online A/B testing on an industrial platform yields \textbf{+2.03\% CTR}, \textbf{+1.21\% CVR}, and \textbf{+1.35\% GMV}, with gains scaling monotonically with user history length. We hope SIF inspires the community to treat historical interactions as first-class, context-rich objects---unlocking sample-level, time-varying signals that static item embeddings fundamentally cannot capture.

\bibliographystyle{ACM-Reference-Format}
\bibliography{sif_paper}

\end{document}